\documentclass{emulateapj}
\usepackage{epsfig,natbib}
\usepackage{epstopdf}
\usepackage{graphicx}
\newcommand{\ms}{\mbox{m\,s$^{-1}~$}}

\newcommand{\ks}{\mbox{km\,s$^{-1}~$}}
\newcommand{\kse}{\mbox{km\,s$^{-1}$}}
\newcommand{\mse}{\mbox{m\,s$^{-1}$}}
\newcommand{\msy}{\mbox{m\,s$^{-1}$\,yr$^{-1}~$}}
\newcommand{\msye}{\mbox{m\,s$^{-1}$\,yr$^{-1}$}}

\newcommand{\msun}{M$_{\odot}~$}
\newcommand{\msune}{M$_{\odot}$}

\newcommand{\rsune}{R$_{\odot}$}

\newcommand{\lsune}{L$_{\odot}$}

\newcommand{\mearth}{$M_\earth$~}
\newcommand{\mearthe}{$M_\earth$}

\newcommand{\rearthe}{$R_\earth$}

\newcommand{\mstar}{\ensuremath{M_{\star}}}
\newcommand{\rstar}{\ensuremath{R_{\star}}}
\newcommand{\lstar}{\ensuremath{L_{\star}}}
\newcommand{\feh}{\ensuremath{[\mbox{Fe}/\mbox{H}]}}
\newcommand{\meh}{\ensuremath{[\mbox{M}/\mbox{H}]}}
\newcommand{\rphk}{\ensuremath{R'_{\mbox{\scriptsize HK}}}}
\newcommand{\lrphk}{\ensuremath{\log{\rphk}}}
\newcommand{\caii}{\ion{Ca}{2} H \& K}

\newcommand{\msini}{\ensuremath{M \sin i}}

\newcommand{\teff}{\ensuremath{T_{\rm eff}}}

\newcommand{\vsini}{\ensuremath{v \sin{i}}}
\newcommand{\plNobs}{\ensuremath{117}}      
\newcommand{\plNobsPre}{\ensuremath{20}}      
\newcommand{\plNobsPost}{\ensuremath{97}}      
\newcommand{\plRMS}{\ensuremath{3.21}}      
\newcommand{\plPerCirc}{\ensuremath{11.4433\pm0.0016}}             
\newcommand{\plTcCirc}{\ensuremath{15874.95\pm0.21}}      
\newcommand{\plKCirc}{\ensuremath{2.94\pm0.28}}      
\newcommand{\plGammaCirc}{\ensuremath{2.47\pm0.88}}      
\newcommand{\plDvdtCirc}{\ensuremath{-0.26\pm0.09}}      
\newcommand{\plJitterPreCirc}{\ensuremath{2.56^{+0.61}_{-0.48}}}      
\newcommand{\plJitterPostCirc}{\ensuremath{1.88\pm0.16}}      
\newcommand{\plMsiniCirc}{\ensuremath{5.35\pm0.75}}      
\newcommand{\plACirc}{\ensuremath{0.0717\pm0.0034}}      
\newcommand{\plRMSPreCirc}{\ensuremath{2.69}}      
\newcommand{\plRMSPostCirc}{\ensuremath{1.93}}      
\newcommand{\plRmsCircBinned}{\ensuremath{0.62}}      
\newcommand{\plPer}{\ensuremath{11.4433\pm0.0017}}             
\newcommand{\plTc}{\ensuremath{15875.09\pm0.45}}      
\newcommand{\plTp}{\ensuremath{15876.6^{+2.4}_{-4.7}}}      
\newcommand{\plECosOm}{\ensuremath{-0.04\pm 0.11}}      
\newcommand{\plESinOm}{\ensuremath{-0.01\pm 0.09}}      
\newcommand{\plEcc}{\ensuremath{0.12^{+0.08}_{-0.06}}}      
\newcommand{\plEccTwoSigma}{0.26}   
\newcommand{\plOm}{\ensuremath{186^{+92}_{-110}}}      
\newcommand{\plK}{\ensuremath{2.93\pm0.29}}      
\newcommand{\plGamma}{\ensuremath{2.43\pm0.90}}      
\newcommand{\plGammaAbsolute}{\ensuremath{11.82\pm0.11}}      
\newcommand{\plDvdt}{\ensuremath{-0.27\pm0.09}}      
\newcommand{\plJitterPre}{\ensuremath{2.54^{+0.62}_{-0.48}}}      
\newcommand{\plJitterPost}{\ensuremath{1.91\pm0.16}}      
\newcommand{\plMsini}{\ensuremath{5.28\pm0.75}}      
\newcommand{\plA}{\ensuremath{0.0717\pm0.0036}}      
\newcommand{\plRMSPre}{\ensuremath{2.70}}      
\newcommand{\plRMSPost}{\ensuremath{1.93}}      
\newcommand{\plName}{Gl\,15\,A}             
\newcommand{\plSpType}{M2\,V}    
\newcommand{\plBV}{\ensuremath{1.55}}      
\newcommand{\plVmag}{\ensuremath{8.08}}      
\newcommand{\plJmag}{\ensuremath{4.82}}      
\newcommand{\plHmag}{\ensuremath{4.25}}      
\newcommand{\plKmag}{\ensuremath{4.03}}      
\newcommand{\plDist}{\ensuremath{3.587\pm0.010}}      
\newcommand{\plMeh}{\ensuremath{-0.22 \pm 0.12}}    
\newcommand{\plFeh}{\ensuremath{-0.32 \pm 0.17}}    
\newcommand{\plTeff}{\ensuremath{3567 \pm 11}}      
\newcommand{\plLogg}{\ensuremath{4.90 \pm 0.17}}      
\newcommand{\plLstar}{\ensuremath{0.02173 \pm 0.00021}}      
\newcommand{\plMstar}{\ensuremath{0.375 \pm 0.057}}   
\newcommand{\plRstar}{\ensuremath{\mathrm{0.3863}\pm\mathrm{0.0021}}}      
\newcommand{\plSval}{\ensuremath{0.527 \pm 0.038}}      

\shortauthors{Howard {et~al.}}
\shorttitle{A Close-in Super-Earth}
\submitted{Astrophysical Journal (accepted)}
\begin{document}
\pagenumbering{arabic}


\title{The NASA-UC-UH Eta-Earth Program: \\
         IV. A Low-mass Planet Orbiting an M Dwarf 3.6 PC from Earth\altaffilmark{1}}
\author{
Andrew W.\ Howard\altaffilmark{2}, 
Geoffrey W.\ Marcy\altaffilmark{3}, 
Debra A.\ Fischer\altaffilmark{4}, 
Howard Isaacson\altaffilmark{3}, 
Philip S.\ Muirhead\altaffilmark{5,6}, 
Gregory W.\ Henry\altaffilmark{7},
Tabetha S.\ Boyajian\altaffilmark{4},
Kaspar von Braun\altaffilmark{8,9}, 
Juliette C.\ Becker\altaffilmark{5},  
Jason T.\ Wright\altaffilmark{10,11}, 
John Asher Johnson\altaffilmark{12,13} 
}
\altaffiltext{1}{Based on observations obtained at the W.\,M.\,Keck Observatory, 
                      which is operated jointly by the University of California and the 
                      California Institute of Technology.  Keck time was granted for this project by the University of Hawaii,  
                      the University of California, and NASA.} 
\altaffiltext{2}{Institute for Astronomy, University of Hawaii, 2680 Woodlawn Drive, Honolulu, HI 96822, USA}
\altaffiltext{3}{Department of Astronomy, University of California, Berkeley, CA 94720-3411, USA} 
\altaffiltext{4}{Department of Astronomy, Yale University, New Haven, CT 06511, USA}
\altaffiltext{5}{Department of Astrophysics, California Institute of Technology, MC 249-17, Pasadena, CA 91125, USA}
\altaffiltext{6}{Current address: Department of Astronomy, Boston University, 725 Commonwealth Ave, Boston, MA  02215, USA}
\altaffiltext{7}{Center of Excellence in Information Systems, Tennessee State University, 3500 John A.\ Merritt Blvd., Box 9501, Nashville, TN 37209, USA}
\altaffiltext{8}{NASA Exoplanet Science Institute, California Institute of Technology, Pasadena, CA 91125, USA}
\altaffiltext{9}{Max-Planck-Institute for Astronomy, K{\"o}nigstuhl 17, 69117 Heidelberg, Germany}
\altaffiltext{10}{Department of Astronomy and Astrophysics, The Pennsylvania State University, University Park, PA 16802, USA}
\altaffiltext{11}{Center for Exoplanets and Habitable Worlds, The Pennsylvania State University, University Park, PA 16802, USA}
\altaffiltext{12}{Center for Planetary Astronomy, California Institute of Technology, 1200 E. California Blvd, Pasadena, CA 91125, USA}
\altaffiltext{13}{Current address: Harvard-Smithsonian Center for Astrophysics, 60 Garden St., Cambridge, MA 02138, USA}

\begin{abstract}
We report the discovery of a low-mass planet orbiting Gl\,15\,A  
based on radial velocities from the Eta-Earth Survey using HIRES at Keck Observatory.
Gl\,15\,Ab is a planet with minimum mass $\msini = \plMsiniCirc$\,\mearthe,
orbital period $P = \plPerCirc$\,d, and an orbit that is consistent with circular.   
We characterize the host star using a variety of techniques.  
Photometric observations at Fairborn Observatory show no evidence for rotational modulation of spots at the 
orbital period to a limit of $\sim$0.1 mmag, thus supporting the existence of the planet.
We detect a second RV signal with a period of 44 days that we attribute to rotational modulation of stellar surface features, 
as confirmed by optical photometry and the \caii\ activity indicator.  
Using infrared spectroscopy from Palomar-TripleSpec, 
we measure an \plSpType\ spectral type 
and a sub-solar metallicity ($\meh = -0.22$, $\feh = -0.32$).
We measure a stellar radius of \plRstar $R_{\sun}$ based on interferometry from CHARA.
\end{abstract}

\keywords{planetary systems --- stars: individual (Gliese 15 A) --- techniques: radial velocity}

\section{Introduction}
\label{sec:intro}

The nearest and brightest stars are among the best studied and hold a special place in the popular imagination.    
The discovery of planets orbiting these stars tells us that the solar neighborhood is potentially rich 
with solar systems.  
Within 7 pc, we know of gas giant planets orbiting 
$\epsilon$~Eridani \citep{Hatzes2000},  
Gl~876 \citep{Marcy1998,Delfosse1998_876}, and
Gl~832 \citep{Bailey2009}, 
intermediate mass planets commonly called Neptunes and super-Earths orbiting 
Gl~674 \citep{Bonfils2007}, 
Gl~876 \citep{Rivera2010}, 
HD~20794 \citep{Pepe2011}, and 
Gl 581 \citep{Mayor2009},
and an approximately Earth-mass (\mearthe) planet orbiting 
$\alpha$~Centauri~B \citep{Dumusque2012}.
To this list we add a 5\,\mearth planet orbiting the star \plName, 
which at 3.6~pc is a member of the sixteenth closest stellar system cataloged by RECONS\footnote{http://www.chara.gsu.edu/RECONS/} 
\citep[e.g.][]{Jao2005,Henry2006}.

This new planet was discovered as part of the Eta-Earth Survey, a census of planets orbiting the nearest stars.
Our target list is composed of 232 G, K, and M stars suitable for high-precision Doppler observations at the Keck Observatory.  
These stars are nearby (within 25 pc), bright ($V < 11$), and have low chromospheric activity ($\lrphk < -4.7$).
Each star is searched for planets ---  particularly close-in, low-mass planets --- 
using a minimum of twenty radial velocity (RV) measurements of $\sim$1\,\ms 
precision from the HIRES spectrometer.  
This survey has detected several low-mass planets \citep{Howard2009_7924, Howard2011_156668, Howard2011_97658} 
and showed that for G and K dwarfs the planet mass function rises steeply with decreasing planet mass: 
small planets are common \citep{Howard2010_science}.
The 66 M dwarfs in the Eta-Earth Survey sample were excluded from that statistical study 
of planet occurrence because many stars lacked sufficient measurements to confidently detect or 
exclude the presence of low-mass planets.  

In this paper, we characterize the planet host star \plName\ 
using a variety of observational techniques (Sec.\ \ref{sec:props}), 
describe Doppler measurements of the star (Sec.\ \ref{sec:keck}), 
announce the existence of the close-in, super-Earth \plName b (Sec.\ \ref{sec:model}), 
and discuss this new planet in the context of the properties of known small planets (Sec.\ \ref{sec:discussion}).

\section{Stellar Characterization}
\label{sec:props}

\plName\ (also known as HD\,1326\,A,  HIP\,1475,  GX Andromedae, and Groombridge 34) 
is a nearby, cool dwarf of type M1 \citep{Reid1995} or M2 (this work). 
The other member of this binary star system, Gl\,15\,B, is fainter and has a spectral type of M3.5 dwarf  \citep{Reid1995}.
\citet{Lippincott1972} measured a small astrometric segment of their orbit, 
giving an AB separation of 146 AU and an orbital period of 2600 years.  
Assuming an edge-on viewing geometry of AB and a circular orbit, 
the maximum RV acceleration of A due to B is $\sim$2 \msye.
Based on an imaging search for companions at 10\, $\mu$m with MIRLIN at Palomar, 
\citet{vanBuren1998} ruled out additional companions to A having projected separations of 9--36 AU 
with \teff\ $>$ 1800 K (\mstar\ $>$ 0.084 \msune). 
\citet{Gautier2007} found no infrared excess for \plName\ at 24, 70 or 160 $\mu m$.

Since \plName\ is a bright, nearby star, we undertook a significant campaign to 
characterize it using a combination of high-resolution optical spectroscopy, 
near infrared (IR) spectroscopy, long-baseline optical/infrared interferometry, 
and high-cadence, broad-band optical photometry.
In the subsections that follow we describe these measurements, 
which are summarized in Table \ref{tab:stellar_params}.

\begin{deluxetable*}{lcc}
\tabletypesize{\footnotesize}
\tablecaption{Adopted Stellar Properties of Gl\,15\,A
\label{tab:stellar_params}}
\tablewidth{0pt}
\tablehead{
  \colhead{Parameter}   & 
  \colhead{Gl\,15\,A} &
  \colhead{Source} 
}
\startdata
Spectral type ~~~~~~& \plSpType & TripleSpec spectra and IRTF Spectral Library \\
$B-V$ (mag) & \plBV &  \citet{Leggett1992} \\
$V$ (mag)   & \plVmag &  \citet{Leggett1992} \\
$J$  (mag)  & \plJmag & \citet{Leggett1992} \\
$H$  (mag)  & \plHmag  & \citet{Leggett1992} \\
$K$ (mag)   & \plKmag &  \citet{Leggett1992} \\
Distance (pc) & \plDist & \cite{vanLeeuwen07} \\
$T_\mathrm{eff}$ (K) &  \plTeff & SED fit \\
log\,$g$ (cgs) & \plLogg & computed from \mstar\ and \rstar \\
\feh\ (dex) & \plFeh$^\mathrm{a}$  &TripleSpec spectra calibrated by \citet{Rojas2012} \\
\meh\ (dex) & \plMeh$^\mathrm{a}$  &TripleSpec spectra calibrated by \citet{Rojas2012} \\
$v$\,sin\,$i$ (km\,s$^{-1}$) &  1.45 $\pm$ 0.6 & \citet{Houdebine2010}\\
$L_{\star}$ ($L_{\sun}$) & \plLstar  & CHARA interferometry and SED fit \\
\mstar\ ($M_{\sun}$) & \plMstar  & TripleSpec parameters with Dartmouth isochrones\\
\rstar\ ($R_{\sun}$) & \plRstar &  CHARA interferometry \\
$S_\mathrm{HK}$  & \plSval  &  \caii\ (HIRES) \\
\enddata
\tablenotetext{a}{The uncertainties for \meh\ and \feh\ from TripleSpec are dominated by systematic errors.  
For comparison with other stars using the same calibration, the photon-limited 
errors on these measurements are 0.02 dex.}
\end{deluxetable*}

\subsection{HIRES Optical Spectroscopy}
\label{sec:hires}

We observed Gl\,15\,A with the HIRES echelle spectrometer \citep{Vogt94} 
on the 10-m Keck I telescope using standard procedures.  
The observations reported here span fifteen years (1997 January through 2011 December)
and were made with an iodine cell mounted directly in front of the 
spectrometer entrance slit to measure precise, relative radial velocities (RVs).
The dense set of molecular absorption lines imprinted 
on the stellar spectra provide a robust wavelength fiducial 
against which Doppler shifts are measured, 
as well as strong constraints on the shape of the spectrometer instrumental profile at 
the time of each observation \citep{Marcy92,Valenti95}.

With the iodine cell removed, we also measured a ``template'' spectrum of \plName\ 
that was used in the Doppler analysis (Sec.\ \ref{sec:keck}).  
For  stars with $\teff \gtrsim 4800$ K, we typically measure stellar parameters using the 
Spectroscopy Made Easy \citep[SME; ][]{Valenti96,Valenti05} LTE spectral synthesis tool. 
However, below this temperature SME is unreliable.
For \plName, we measured stellar parameters using the techniques below.

Measurements of the cores of the \caii\ lines of each HIRES spectrum 
(outside the $\sim$5000-6200~\AA\ region affected by iodine lines) show 
modest levels of chromospheric activity, as quantified by the $S_{\mathrm{HK}}$ and \lrphk\ indices \citep{Isaacson2010}.  
The $S_{\mathrm{HK}}$ values (Table \ref{tab:keck_vels_gl15a}) are computed to a precision of 0.001 and 
carry single measurement uncertainties of $\sim$0.002.  Values in Table \ref{tab:keck_vels_gl15a} that are reported to
a precision less than 0.001 are the result of binning multiple measurements in a 2 hour timespan.
These measurements  are variable on both short and long timescales.  
Over our 15 year baseline, we detect a $9 \pm 2.5$ year cycle with a semi-amplitude of $\sim$0.05 
(in the dimensionless units of $S_{\mathrm{HK}}$), which is a $\sim$10\% fractional change.
This variation may be a magnetic activity cycle analogous to the solar cycle.  
We are unable to check for a similar variation in the optical photometry from Fairborn Observatory below 
(Sec.\ \ref{sec:photometry}), which span only four years.

\begin{figure}
\plotone{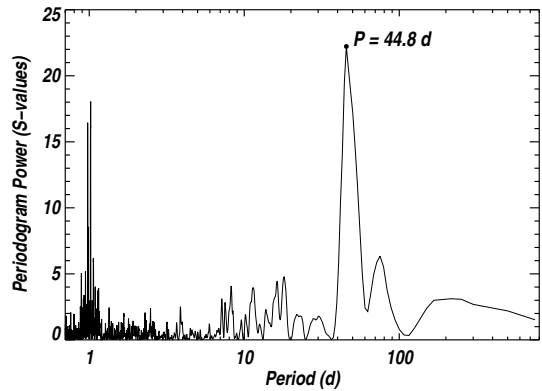}
\caption{Lomb-Scargle periodogram of S$_{\mathrm{HK}}$ values from  59 measurements during 
               the 2011 observing season (BJD $>$ 2,455,500).  
               We interpret the prominent peak near 44 days as due to modulation of stellar surface 
               features (spots and plagues) at the stellar rotation period.  
               The peaks near 1 day are aliases of our observing cadence and longer period signals.
               }
\label{fig:pergram_sval}
\end{figure}

To measure variability on shorter timescales, we examined 59 $S_{\mathrm{HK}}$ measurements 
from the 2011 observing season (BJD $>$ 2,455,500).  
A Lomb-Scargle periodogram of these data (Figure \ref{fig:pergram_sval}) shows a clear periodicity 
near 44 days that we interpret as the rotation period.  
As shown below, we also detect $\sim$44 day modulations in optical photometry and the RV time series, 
which is consistent with the rotational modulation of stellar surface features.

\subsection{Palomar-TripleSpec Infrared Spectrscopy}
\label{sec:triplespec}

We observed \plName\ with the TripleSpec Spectrograph on the Palomar Observatory 
200-inch Hale Telescope on UT 12 February 2012.   
TripleSpec is a near-infrared, long-slit spectrograph covering 1.0--2.5 $\mu$m simultaneously 
with a resolving power of 2700 \citep{Wilson2004, Herter2008}.  
The TripleSpec detector does not have a shutter, and instead records differences between multiple 
non-destructive readouts for individual exposures \citep{Fowler1990}.  
The readout time sets the minimum exposure time of the detector to 3.8 seconds.  
To prevent saturation of this bright target, we guided with only a wing of the seeing-limited image 
of \plName\  on the slit.  
Two positions along the slit (A and B) were used.  Exposures were taken in an ABBA pattern 
with the minimum exposure time.

\begin{figure}
\plotone{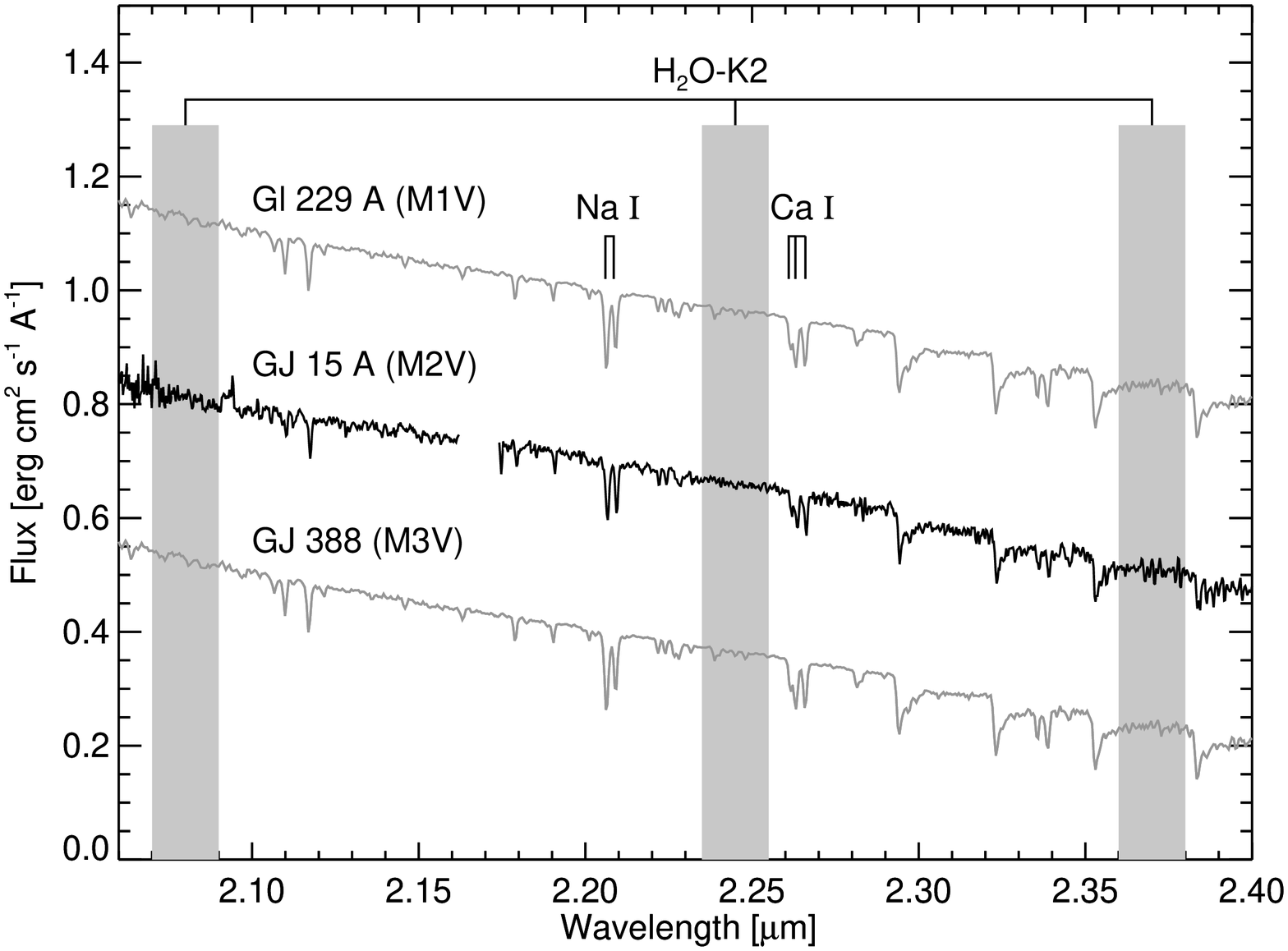}
\caption{Palomar-TripleSpec $K$-band spectrum of \plName, with comparison stars drawn from the 
IRTF Spectral Library \citep[KHM spectral types][]{Cushing2005, Rayner2009}. 
The templates are adjusted to the same scale as the \plName\ spectrum using a ratio of the median flux in K band, 
and then artificially offset.  
We used spectral indices defined in \citet{Rojas2012} to measure the 
\teff, \meh, and \feh :  the $\rm H_2O$-K2 index (gray shaded regions) and 
the equivalent widths of the \ion{Na}{1} doublet and \ion{Ca}{1} triplet features (indicated).}
\label{fig:tspec}
\end{figure}

We reduced the data using the SpexTool program, 
modified for Palomar TripleSpec \citep[][M. Cushing, private communication]{Cushing2004}.
The spectra were cleansed of telluric absorption lines by comparison with spectra of 
a nearby A0V star (BD+43 61) and were flux calibrated using the {\tt xtellcor} package \citep{Vacca2003}.  
The brightness of \plName\ resulted in a discontinuity in the reduced spectrum at the location 
of a quadrant boundary in the detector.  
We manually adjusted the normalization of the spectrum to remove the discontinuity.  
Figure \ref{fig:tspec} shows the reduced $K$-band spectrum of \plName, 
with comparison spectra of bracketing spectral types.
 
Moderate resolution $K$-band spectra of M dwarfs are sensitive to 
stellar temperature \citep{Covey2010} and metallicity \citep{Rojas2010, Rojas2012}.  
For \plName , we measured spectral indices and used the \citet{Rojas2012} calibrations to estimate \teff, \meh, and \feh.
We measured the equivalent widths of 3.430  $\pm$ 0.098 \AA\ and 3.148 $\pm$ 0.103 \AA\ for the
$K$-band \ion{Na}{1} and \ion{Ca}{1} absorption features, respectively.  
For the $\rm H_2O$--K2 index, we measured a value of 0.95434 $\pm$ 0.0030.
From these measurements we infer 
\feh\ =  $-$0.32 $\pm$ 0.17 and \meh\ = $-$0.22 $\pm$ 0.12 using the \citet{Rojas2012} calibration.  
These metallicity estimates are consistent with the broadband photometric calibration 
by \citet{Johnson2009}, which predicts \feh\ = $-$0.25 $\pm$ 0.07.

Following \citet{Rojas2012}, we interpolated \meh\ and the $\rm H_2O$--K2 index on a surface of 
\meh, $\rm H_2O$--K2, and \teff\  calculated using synthetic spectra \citep{Allard2011}.  
The resulting temperature \teff\ = 3568  $\pm$ 52 K is consistent with the temperature estimate of 3567 $\pm$ 11 K 
from fitting the spectral energy distribution and CHARA interferometry (Sec.\ \ref{sec:chara}).

We estimated uncertainties using a Monte Carlo approach combined with calibration 
uncertainties from \citet{Rojas2012}.  SpexTool reports per-channel uncertainties based on the 
photon noise and read noise in both the target and calibration data.  We created 1000 instances of the data, 
each with random, normally-distributed noise added based on the reported per-channel uncertainties.  
For each instance, we calculated the spectral indices and report the standard deviation in the distribution of 
values as the uncertainty.  We also calculated \teff, \meh, and \feh\ for each iteration, 
however these are also subject to uncertainties in the calibration relations.  
\citet{Rojas2012} calculates an uncertainty of 0.17 in the \feh\ relation and 0.12 in the \meh\ relation, 
which dominate over the Monte Carlo uncertainty estimations.  
We estimate the uncertainty in the effective temperature calculation to be 50 K, 
which also dominates over the Monte Carlo estimation.

We interpolated the measured \meh\ and \teff\ on stellar evolutionary models to 
determine \mstar, \rstar, and \lstar.  
We explored 5-Gyr isochrones from two sets of evolutionary models: 
the Dartmouth evolutionary models \citep[e.g.][]{Dotter2008, Feiden2011} and 
the BCAH evolutionary models \citep{Baraffe1998}.  
The age assumption does not change stellar parameters by more 1\% for ages over 1 Gyr.  
To estimate uncertainties, we again used a Monte Carlo approach, 
interpolating 1000 iterations of \teff\ and \meh\ onto the isochrones, with 
each iteration containing normally-distributed noise based on the calculations described above.  
For the Dartmouth models respectively, we estimate 
\mstar\ = 0.375 $\pm$ 0.057 \msune, 
\rstar\ = 0.363 $\pm$ 0.052 \rsune, and 
\lstar\ = 0.0191 $\pm$ 0.0040 \lsune.
For the BCAH models respectively, we estimate 
\mstar\ = 0.358 $\pm$ 0.065 \msune, 
\rstar\ = 0.342 $\pm$ 0.057 \rsune, and 
\lstar\ = 0.0170 $\pm$ 0.0045 \lsune.
These radius and luminosity estimates are consistent with those from 
CHARA interferometry and spectral energy distribution (SED) modeling (Sec.\ \ref{sec:chara}).

For comparison, we estimate the mass of \plName\ using the absolute $K$-band mass-luminosity calibration of \cite{Delfosse2000}.
We find $M_\star = 0.404 \pm 0.033$ \msun based on $K = \plKmag$ mag \citep{Leggett1992} and a distance of \plDist\ pc 
\citep{vanLeeuwen07}.  The mass uncertainty comes from 0.18 mag rms errors in fitting the $K$-band photometry in 
\cite{Delfosse2000} (see Sec.\ 3.6 of \citet{Johnson2012}).

\subsection{CHARA Interferometry}
\label{sec:chara}

\subsubsection{Interferometry and Stellar Diameter}

We measured the angular diameter of \plName\ using visibility interferometry.  We observed
\plName\ over seven nights between 2008 Sept and 2011 August using the 
Georgia State University Center for High Angular Resolution Astronomy (CHARA) Array \citep{2005tenBrummelaar} 
as part of a survey of K and M dwarfs \citep{Boyajian2012}.  
We obtained forty interferometric observations in $H$ and $K$ bands in single baseline mode, 
using six of CHARA's longest available baselines ($217 < B_{\rm max} < 331 $ m). 
Our observational strategy is described in \citet{2011vonBraun_b,2011vonBraun}, 
including the use of calibrator stars to remove the influence of atmospheric and instrumental systematics. 
Our calibrator stars were chosen to be near-point-like sources of similar brightness as \plName\ 
and are separated by small angular distances.  These stars are HD 6920, HD 905, and HD 3765.

We measured a limb-darkening-corrected \citep{Claret2000} angular diameter of 
1.005 $\pm$ 0.005 milliarcseconds (mas). 
Using the  \citet{vanLeeuwen07} distance, we convert the angular diameter to a physical diameter of \rstar\ = $0.3846 \pm 0.0023$ \rsune, 
in agreement with previous CHARA measurements of $0.379\pm0.006$ \rstar\  
\citep{Berger2006} and $0.393 \pm0.023$ \rstar\ \citep{2009vanBelle}.

With this well-measured stellar radius and rotation period, we can make a comparison with the spectroscopic broadening kernel, \vsini.  
For an equatorial viewing geometry ($\sin i$ = 1), we the expect 
\vsini\ $\approx V_{\mathrm{rot}} \approx 2\pi$\rstar/$P_{\mathrm{rot}}$.  
For \plName, we find $2\pi$\rstar/$P_{\mathrm{rot}}$= 0.44 \kse, 
which is inconsistent at the 1.7-$\sigma$ level with the measured \vsini\ = 1.45 $\pm$ 0.6 \ks \citep{Houdebine2010}.

\subsubsection{Spectral Energy Distribution}
\label{sec:sed}

\begin{figure}
\epsscale{1.15}
\plotone{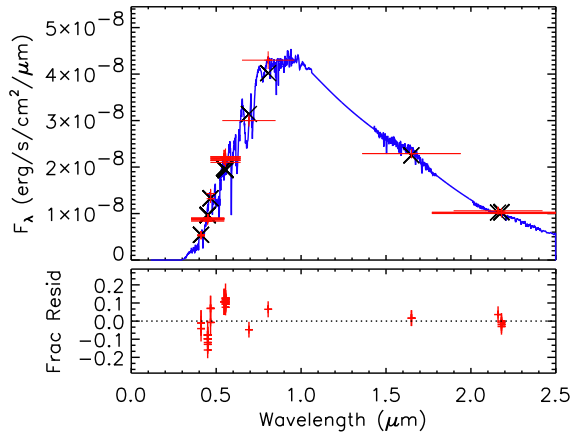}
\caption{
SED fit for \plName\  with a M2V spectral template (blue) from \citet{Pickles1998}  
and photometry (red) from the literature.
Filter bandpasses are indicated by the horizontal lengths of the red photometric measurements.   
The black ``X'' symbols show the template flux integrated over the filter bandpasses.  
Residuals between these integrated model and measured fluxes are in the lower panel.
For details, see \S \ref{sec:sed}}
\label{fig:sed}
\end{figure}

Similar to the procedure outlined in \citet{2012vonBraun}, we fitted the SED (Figure \ref{fig:sed}) based on the 
spectroscopic templates of \citet{Pickles1998} to literature photometry published in 
\citet{Cowley1967, Johnson1953, Upgren1974, Niconov1957, Argue1966, Johnson1965, Erro1971, Mermilliod1986, Hauck1998, Twarog1980, Olsen1993, Olson1974, Cutri2003} 
and cataloged in \citet{Gezari1999}. 
We did not include interstellar extinction from the SED fit due to the proximity of \plName.

The SED fit produces a bolometric flux of $F_{\mathrm{bol}} = (5.42  \pm 0.04) \times 10^{-8}$ erg\,cm$^{-2}$\,s$^{-1}$. 
The $\sim$1\% fractional error accounts only for photometric and fitting uncertainties 
and does not include systematic uncertainties (e.g., see Sec.\ 2.2 of \citet{vonBraun2014}).
Combining $F_{\mathrm{bol}}$ with the distance, we derive a luminosity of $L$ = 0.02173 $\pm$ 0.00021~$L_{\odot}$.  
We derive an effective temperature of  3563 $\pm$ 11 K using the modified 
Stefan-Boltzmann Law 
$\teff ({\rm K}) = 2341 (F_{\rm  bol}/\theta_{\rm LD}^2)^{\frac{1}{4}}$ from \citet{2012vonBraun}.

Using the approach in \citet{2011vonBraun_b} and based on the equations of \citet{Jones2010} 
we calculate the inner and outer boundaries of the habitable zone around \plName\ 
to be 0.14 and 0.29 AU, respectively.
With a semi-major axis of 0.074 AU (see Sec.\ \ref{sec:model}), \plName b orbits interior to the habitable zone.

\subsection{Photometry from Fairborn Observatory}
\label{sec:photometry}

We measured the brightness of \plName\ over four observing seasons with the T12 0.80~m 
automatic photometric telescope (APT), one of several automatic telescopes 
operated by Tennessee State University at Fairborn Observatory \citep{Eaton2003}.  
The APTs can detect short-term, low-amplitude brightness 
changes in solar-type stars resulting from rotational modulation in the 
visibility of active regions, such as starspots and plages 
\citep[e.g.,][]{Henry1995} and can also detect longer-term variations 
produced by the growth and decay of individual active regions and the 
occurrence of stellar magnetic cycles \citep[e.g.,][]{Henry1995b,Hall2009}.  
The TSU APTs can disprove the hypothesis that RV variations
are caused by stellar activity, rather than planetary reflex motion
\citep[e.g.,][]{Henry2000}.  Several cases of apparent periodic RV  
variations in solar-type stars induced by the presence of photospheric 
starspots have been discussed in the literature 
\citep[e.g.,][]{Queloz2001,Paulson2004,Bonfils07,Forveille09}.  
Photometry of planetary candidate host stars is also useful to search for 
transits of the planetary companions 
\citep[e.g.,][]{Henry2000b,Sato05,Gillon2007,Barbieri2007}.

The T12 0.80~m APT is equipped with a two-channel photometer that uses two 
EMI 9124QB bi-alkali photomultiplier tubes (PMTs) to make simultaneous 
measurements of one star in the Str\"omgren $b$ and $y$ passbands.  
We report differential measurements of the target star with respect to  
three comparison stars: 
HD\,571 ($V= 5.01$, $B-V =  0.41$),  
HD\,818 ($V= 6.63$, $B-V =  0.40$), and 
HD\,1952 ($V= 6.66$, $B-V =  0.41$).  
The T12  APT is functionally identical to the T8 APT described in \citet{Henry1999}.
All photometric measurements were
made through a 45\arcsec\ focal-plane diaphragm, thus excluding the
light from the two 11th magnitude visual companions Gl~15 B and C.
The observing sequence and conditions for rejecting photometry in non-photometric conditions 
are described in \citet{Henry1999}.

\begin{figure}
\plotone{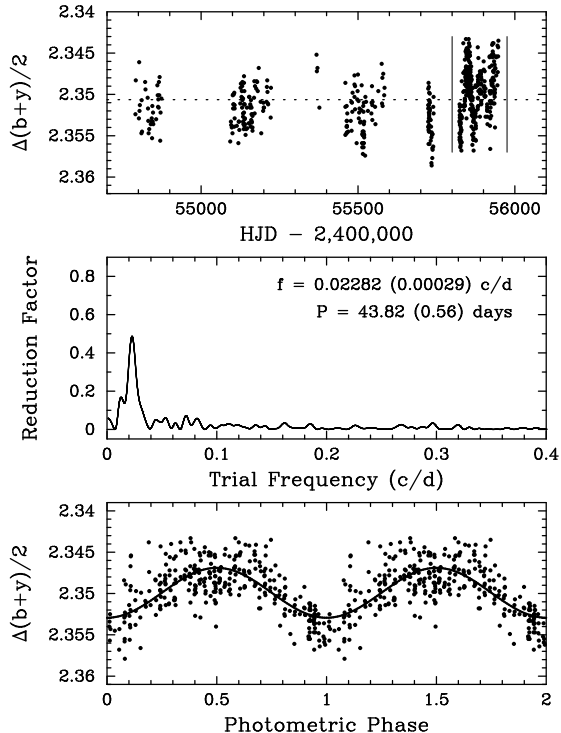}
\caption{Top panel:  Stro\"mgren $(b+y)/2$ differential 
magnitudes of \plName\ plotted against heliocentric Julian Date for four observing seasons (2008-2011).  
The standard deviation of these normalized observations from their mean (dotted line) is 3.1\,mmag.  
The vertical lines in the top panel set off the best covered and most coherent segment of the light curve 
that shows rotational modulation of spots.  The two lower panels focus on this high cadence subset of the photometry.
Middle panel: Frequency spectrum of the high cadence photometry showing a peak corresponding to 
a period of $\sim44$ days, which we interpret as the stellar rotation period.
Bottom panel: Photometry from the high cadence interval phased to a 44 day period.  
Two cycles are shown (with data repeated) along with a least-squared sine function fit having 
a peak-to-peak amplitude of 6 mmag.
}
\label{fig:apt_photometry_3panel}
\end{figure}

During four consecutive observing seasons starting in 2008, 
the APT acquired 578 differential brightness measurements of \plName.
We combined the $b$ and $y$ differential magnitudes into $(b+y)/2$ measurements, 
achieving typical single measurement precision of  1.5--2.0\,mmag \citep{Henry1999}.
These measurements are plotted in the top panel of Figure \ref{fig:apt_photometry_3panel} 
and have a standard deviation of 3.1\,mmag, which is somewhat larger than measurement uncertainties.

In the 2011 observing season, we increased the  cadence to several observations per night. 
A periodogram of these measurements (Figure \ref{fig:apt_photometry_3panel}, middle panel) 
shows Fourier power at a period of 43.82 $\pm$ 0.56 days.  The phased photometry from this season
is shown in the bottom panel of Figure \ref{fig:apt_photometry_3panel}.
The APT photometry show no evidence for rotational modulation of spots at the 
orbital period to a limit of $\sim$0.1 mmag, thus supporting the interpretation that the 11.44\,d RV signal is 
due to an orbiting planet (Sec.\ \ref{sec:APT_conf}).

\section{Keck-HIRES Doppler Measurements}
\label{sec:keck}

We measured the Doppler shift of each star-times-iodine HIRES spectrum (Sec.\ \ref{sec:hires}) using a 
modeling procedure descended from \citet{Butler96b} and described in \citet{Howard2011_156668}.  
The velocity and corresponding uncertainty for each
observation is based on separate measurements for $\sim$700 spectral chunks each 2 \AA\ wide.
Once the  planet announced here emerged as a candidate in October 2010, 
we increased the nightly cadence to three consecutive observations per night to reduce the 
Poisson noise from photon statistics.  
We calculated mean velocities for multiple observations in 2\,hr intervals.  
The RMS of these measurements is \plRMS\ \mse.  
The relative RVs and simultaneous $S_\mathrm{HK}$ values  are listed in Table \ref{tab:keck_vels_gl15a}.  
The absolute RV of \plName\ relative to the solar system barycenter is \plGammaAbsolute\ \ks \citep{Chubak2013}.
The RVs in Table \ref{tab:keck_vels_gl15a} are corrected for motion of Keck Observatory through the 
Solar System (barycentric corrections) and 
secular acceleration, but not for any measured or assumed motion of \plName\ from interactions with Gl\,15\,B or C.

Measurements made after the HIRES CCD upgrade in 2004 August suffer from smaller systematic errors.  
As described in the Sec.\ \ref{sec:model}, when modeling the measurements we allowed for a zero-point offset between the 
\plNobsPre\ ``pre-upgrade'' and \plNobsPost\ ``post-upgrade'' RVs, as well as differing amounts of jitter.

\section{Planet Detection and Orbital Model}
\label{sec:model}

\begin{figure}
\plotone{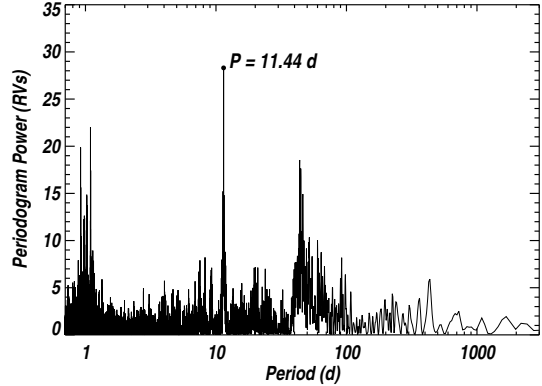}
\caption{Lomb-Scargle periodogram of RV measurements of Gl\,15\,A. 
     The tall peak near $P=11.44$ d suggests a planet with that orbital period.
     }
\label{fig:pergram_gl15a}
\end{figure}

\begin{figure}
\plotone{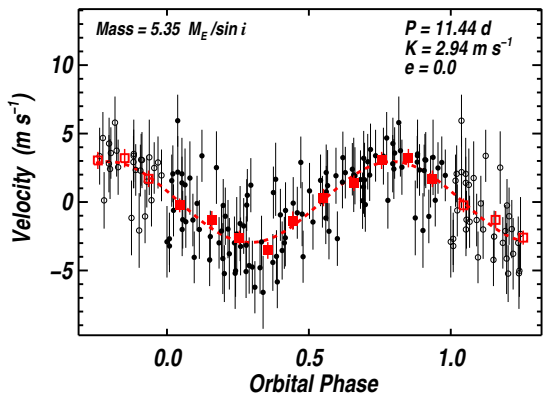}
\caption{Single-planet model for the Keck-HIRES RVs of Gl\,15\,Ab.  
       Filled black circles represent phased measurements  
       while the open black circles represent the same velocities wrapped one orbital phase.
       The error bars show the quadrature sum of measurement uncertainties and jitter.
       Red squares show RVs binned in 0.1 phase increments and have an RMS to the model of 
       \plRmsCircBinned\ \mse.
       The best-fit circular orbital solution is shown as a dashed red line.}
\label{fig:phased_gl15a}
\end{figure}

We modeled the RV time series as a single planet in Keplerian orbit around the star \plName. 
We searched for periodic signals by computing a Lomb-Scargle periodogram 
\citep{Lomb76,Scargle82}  of the RVs (Figure \ref{fig:pergram_gl15a}) and found a dominant peak at 11.44~days.  
We seeded single-planet Keplerian models with that period and a variety of other periods
using the orbit fitting techniques described in \citet{Howard09b} 
and the partially linearized, least-squares fitting procedure described in \citet{Wright09b}.  
Our adopted model (Table \ref{tab:orbital_params_gl15a}) has an orbital period near 11.44 days.
We also include a linear trend of \plDvdtCirc\ \msye, which is consistent with the mass and separation of 
the \plName B system (see Sec.\ \ref{sec:props}).
The best-fit circular, single-planet model is plotted in Figure \ref{fig:phased_gl15a}.

\begin{deluxetable*}{lc}
\tablewidth{0pc}
\tabletypesize{\scriptsize}
\tablecaption{
	Orbital Solutions for Gl\,15\,Ab
	\label{tab:orbital_params_gl15a}
}
\tablehead{
	\colhead{~~~~~~~~Parameter~~~~~~~~}	&
	\colhead{Value}                         
}
\startdata
\noalign{\vskip -3pt}
\sidehead{Circular Orbit Model (adopted)}
~~~~$P$ (days)     & \plPerCirc \\
~~~~$T_c\,^{a}$ (BJD -- 2,440,000) & \plTcCirc \\
~~~~$K$ (m\,s$^{-1}$)       & \plKCirc \\
~~~~$\gamma$ (\mse)        & \plGammaCirc \\
~~~~$dv/dt$ (\msye) & \plDvdtCirc \\
~~~~$M$\,sin\,$i$ ($M_\earth$) & \plMsiniCirc \\
~~~~$a$ (AU)                & \plACirc \\
~~~~RMS to best-fit model, pre-upgrade RVs ($\mse$) & \plRMSPreCirc \\
~~~~RMS to best-fit model, post-upgrade RVs ($\mse$) & \plRMSPostCirc \\
~~~~$\sigma_{\mathrm{jit,pre}}$ ($\mse$) & \plJitterPreCirc \\
~~~~$\sigma_{\mathrm{jit,post}}$ ($\mse$) & \plJitterPostCirc \\
\sidehead{Eccentric Orbit Model}
~~~~$P$ (days)     & \plPer \\
~~~~$T_c\,^{a}$ (BJD -- 2,440,000) & \plTc \\
~~~~$T_p\,^{b}$ (BJD -- 2,440,000) & \plTp \\
~~~~$e \cos \omega $                     & \plECosOm \\
~~~~$e \sin \omega $                     & \plESinOm \\
~~~~$e$                     & \plEcc \\
~~~~$\omega$ (deg)          & \plOm \\
~~~~$K$ (m\,s$^{-1}$)       & \plK \\
~~~~$\gamma$ (\mse) of relative RVs        & \plGamma \\
~~~~$dv/dt$ (\msye) & \plDvdt \\
~~~~$M$\,sin\,$i$ ($M_\earth$) & \plMsini \\
~~~~$a$ (AU)                & \plA \\
~~~~RMS to best-fit model, pre-upgrade RVs ($\mse$) & \plRMSPre \\
~~~~RMS to best-fit model, post-upgrade RVs ($\mse$) & \plRMSPost \\
~~~~$\sigma_{\mathrm{jit,pre}}$ ($\mse$) & \plJitterPre \\
~~~~$\sigma_{\mathrm{jit,post}}$ ($\mse$) & \plJitterPost \\
\enddata
\tablenotetext{a}{Time of inferior conjunction or time of transit if viewed edge-on.}
\tablenotetext{b}{Time of periastron passage.}
\end{deluxetable*}

We estimated orbital parameter uncertainties using a Markov Chain Monte Carlo (MCMC) 
method \citep{Ford2005,Ford2006} with the Metropolis-Hastings algorithm \citep{Metropolis1953,Hastings1970} 
and a Gibbs sampler \citep{Geman1984}.   
We report the median, 84.1\%, and 15.9\% levels of the marginalized posterior parameter distributions 
in Table \ref{tab:orbital_params_gl15a}. 
Instead of minimizing $\chi^2$, we (equivalently) maximized the logarithm of the likelihood, 
\begin{eqnarray}
\ln \mathcal{L} &=& 
- \sum_{i=1}^{N_{\mathrm{pre}}} \frac{(v_i-v_\mathrm{m}(t_i))^2}{2(\sigma_i^2+\sigma_{\mathrm{jit,pre}}^2)}
-  \ln \sqrt{2\pi(\sigma_i^2+\sigma_{\mathrm{jit,pre}}^2)}    \nonumber \\
&& - \sum_{i=1}^{N_{\mathrm{post}}} \frac{(v_i-v_\mathrm{m}(t_i))^2}{2(\sigma_i^2+\sigma_{\mathrm{jit,post}}^2)}
-  \ln \sqrt{2\pi(\sigma_i^2+\sigma_{\mathrm{jit,post}}^2)},  \nonumber \\
\label{eq:likelihood}
\end{eqnarray}
where $v_i$ and $\sigma_i$ are the $i$th velocity measurement and its associated measurement error 
from among the $N_{\mathrm{pre}}$ and $N_{\mathrm{post}}$ measurements acquired before and after the 2004 upgrade of HIRES; 
$v_\mathrm{m}(t_i)$ is the Keplerian model velocity at time $t_i$; 
$\sigma_{\mathrm{jit,pre}}$  and $\sigma_{\mathrm{jit,post}}$  are the jitter estimates of the pre- and post-upgrade data sets.
The first term in each sum represents the usual normalized sum of squared residuals ($\chi^2$). 
Following \cite{Johnson2011_18pl}, we allowed jitter to float in the MCMC analysis, 
as controlled by the second terms in each sum in eq.\ \ref{eq:likelihood}.  
Our model uses separate jitter parameters for the pre- and post-upgrade RV data sets.  
The jitter estimates for these datasets (see Table \ref{tab:orbital_params_gl15a}) 
show the improvement in HIRES measurement precision after the 2004 upgrade.
We adopted a Gregory eccentricity prior \citep{Gregory2010} and non-informative priors 
on other parameters.
The jump parameters of the MCMC model were 
orbital period $P$, 
a time of transit (or inferior conjunction) $T_c$, 
Doppler semi-amplitude $K$, 
Lagrangian parameters $e \cos \omega$ and $e \sin \omega$, 
RV zero-point $\gamma$, 
RV offset between pre- and post-upgrade data sets, 
linear RV trend $dv/dt$, 
and pre- and post-upgrade RV jitter terms $\sigma_{\mathrm{jit,pre}}$ and $\sigma_{\mathrm{jit,post}}$.
From these jump parameters we derived the remaining parameters listed in 
Table \ref{tab:orbital_params_gl15a}.  

We justify the inclusion of a linear RV trend, $dv/dt$, as follows.  
For the circular planet model with a trend as a free parameter (Table \ref{tab:orbital_params_gl15a}, top), 
only 0.3\% of the MCMC trials have $dv/dt > 0$.  
Thus, a negative RV trend is preferred with approximately 3-$\sigma$ significance.
The trend is also physically well motivated by the distant orbit of the star Gl\,15\,B, 
and is consistent with the $\sim$1 \msy expected amplitude (order of magnitude).

We  considered circular and eccentric single-planet models.  
We adopted the circular orbit (Table \ref{tab:orbital_params_gl15a}, top) for two reasons.  
First,  the posterior distributions for 
$e \cos \omega$ and $e \sin \omega$ are consistent with zero to within 1-$\sigma$ in the floating eccentricity model.  
Second, the $\chi^2$ value (not reduced) of the best-fit eccentric model 
is only smaller than the value for the best-fit circular model by 0.3, 
which fails to justify the addition of two model parameters.  
The eccentric model rules out $e > \plEccTwoSigma$ with 95\% confidence.

\subsection{Photometric confirmation and transit search}
\label{sec:APT_conf}

\begin{figure}
\plotone{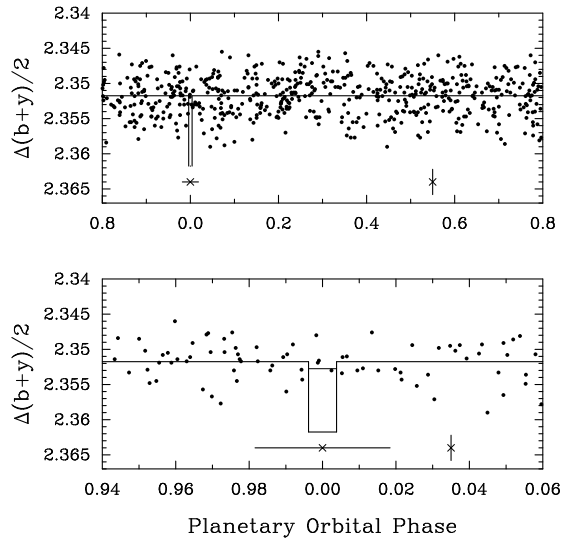}
\caption{APT Photometry from four observing seasons (2008-2011) phased to the best-fit orbital period 
	(Table \ref{tab:orbital_params_gl15a}).  
	The photometric means in the final three observing seasons were adjusted to remove yearly offsets.
	The top panel shows the entire orbital phase, while the bottom panel focusses on photometry near 
	the phase of predicted transits.  
	Photometry in the top panel shows no apparent modulation at the orbital period, 
	which is consistent with our planetary interpretation of the RV modulation.  
	Transit depths of 1 and 10 mmag are shown for a toy model with a transit duration of 2.1 hours for 
	an equatorial transit.  
	The sparse photometry rules out transits if the planet's atmosphere is extremely bloated, 
	but are insensitive to transits if the planet's atmosphere contributes minimally to its radius (see text for details).  
	The vertical bar shows the typical photometric error of 1.8 mmag.  
	The transit time uncertainty for the circular orbit model is indicated by the horizontal line segment 
	centered at phase 0.0.
               }
\label{fig:apt_photometry_phased}
\end{figure}

We searched for periodic variability of the APT photometry (Sec.\ \ref{sec:photometry}) 
at the orbital period of 11.44\,d, but found none.  
This non-detection strengthens the planetary interpretation for the 11.44\,d RV signal.  
A least-squares sine fit to the photometry at the best-fit orbital period gives a semi-amplitude of $0.09 \pm 0.14$\,mmag.   
As shown in the top panel of Figure \ref{fig:apt_photometry_phased}, 
this tight limit supports the hypothesis that the RV signal is due to stellar reflex motion from a planet in motion, rather than spots.

\plName b has a 2\% \textit{a priori} probability of having an orbital inclination $i$ 
that gives rise to eclipses as seen from Earth.   
While this  geometry is unlikely, it is instructive to consider the photometric detectability of transits.
For plausible planets with radii 1.5--4.0 \rearthe, 
corresponding to densities of 8--0.4 g\,cm$^{-3}$, the transit depths are 1.4--10 mmag.
If the planet has an Earth-like density,  
then $R_p$ = 1.8 \rearthe, which gives a transit depth of 2.0 mmag.
Equatorial transits will last 2.1 hr, as shown by the width of the box-shaped transit model 
in the bottom panel Figure \ref{fig:apt_photometry_phased}.
Our sparse APT photometry casts doubt on the presence of a very low density transiting planet.  
However, we refrain from making quantitative statements about the sizes of planets that can be excluded based on the current data.  
To date, only $\sim$4 points are contained in the expected best-fit transit window.
A dedicated photometric campaign from space or the ground could likely detect or exclude all 
of the planet sizes described above.

\subsection{Null hypothesis considered}

We considered the null hypothesis---that the observed RVs are the chance arrangement 
of random velocities masquerading as a coherent signal---by calculating  false alarm probabilities (FAPs) 
and Bayesian Information Criteria (BIC).   
Using the method described in \citet{Howard09b}, 
we computed the improvement in $\Delta\chi^2$ from a constant velocity model to a 
Keplerian model for $10^3$ scrambled data sets.  
We allowed for eccentric single-planet orbital solutions in the scrambled data sets.
We found that no scrambled data set had a larger $\Delta\chi^2$ value than the measured velocities did, 
implying an FAP of $<$~0.001.

As an additional check, we assessed statistical significance by computing 
the BIC \citep{Schwarz1978,Liddle2004}
for the single planet circular model (Table \ref{tab:orbital_params_gl15a}, top) 
and the null hypothesis, a model with an RV trend varying linearly with time.  
The BIC is defined as $\chi^2_{\mathrm{min}} + k \ln N_{\mathrm{obs}}$, 
where $\chi^2_{\mathrm{min}}$ is the unreduced $\chi^2$ value of the best-fitting model 
having $k$ degrees of freedom.  
$N_{\mathrm{obs}}$ = \plNobs\ is the number of RV measurements.  
Differences of 2--6, 6--10, $>$10 between BIC values for the two models indicates 
that there is positive, strong, and very strong evidence for the more complex model \citep{Kass1995}.  
The single planet model (Table \ref{tab:orbital_params_gl15a}) 
is strongly supported over the null hypothesis with $\Delta$BIC = 24.  

\begin{figure}
\plotone{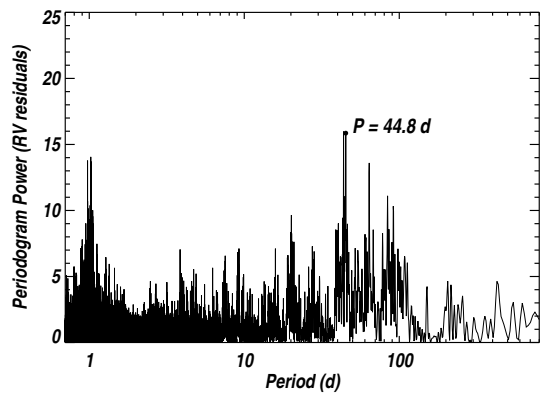}
\caption{Lomb-Scargle periodogram of the RV residuals to the single-planet, circular orbit model for \plName b.  
We interpret the peaks near 44 days as artifacts of stellar rotation because modulation at this period is also 
detected in \caii\ lines and optical photometry.  }
\label{fig:pergram_vres}
\end{figure}

\subsection{A second RV signal}
\label{sec:rv_second_sig}

In addition to the single-planet model presented above, we also considered two-planet Keplerian models.  
The small residuals to the one-planet fit constrain the family of possible additional planets 
to those with small Doppler amplitudes -- low mass and/or distant.  
We computed a periodogram of the RV residuals (Figure \ref{fig:pergram_vres}) to the single-planet fit and found  
several periods with considerable power in the range $\sim$30--120\,d, 
with the two largest peaks near 44\,d.  
These peaks correspond to Doppler signals with $\sim$1--2\,\ms semiamplitudes 
and represent possible second planets with masses in the range $\sim$2--10 \mearthe.  

We considered two-planet orbital solutions with $P_b$ seeded with the best-fit value 
from the single-planet model and $P_c$ seeded with peaks in the residual periodogram.  
Fits with $P_c$ seeded with periods corresponding to either of two tallest peaks in Figure \ref{fig:pergram_vres} 
yield the most statistically significant two-planet fits.  
If real, this signal would represent a 5 \mearth planet in a 0.18\,AU orbit  
within the classically defined habitable zone.  
The signal (at either of the two periods near 44~d) has an amplitude of 1.8~\mse.

However, we do not interpret the 44~d RV signal as a planet.  
The broad periodogram power in the $\sim$30--120\,d range, with two closely spaced peaks near 44\,d, 
suggests that it is due to rotationally-modulated spots.  
Such spots appear and disappear on timescales of weeks and months on a range of differentially rotating stellar latitudes, 
injecting a quasi-coherent noise in the RV time series.  
In contrast, an orbiting planet creates a coherent RV signature.  
The spot interpretation is strengthened with the detection of periodic signals in the 
photometry (Sec.\ \ref{sec:photometry})
and \caii\ line strengths (Sec.\ \ref{sec:hires}), also with $\sim$44\,d periods.
We measured the correlation between the RV residuals to the one-planet fit and 
the simultaneous $S_\mathrm{HK}$ values.  
The Pearson correlation coefficient of $r$ = $+$0.41 demonstrates a modest correlation between these quantities.  
In comparison, the RVs in Table \ref{tab:keck_vels_gl15a} (before subtracting the single-planet fit) 
are less correlated with the $S_\mathrm{HK}$ values, with $r$ = $+$0.24.  
We conclude that the $\sim$44~d RV signal is mostly likely not due to an orbiting planet.  

As an additional check, we computed  ``running periodograms'' \citep{Howard2011_156668} 
of the 11.4\,d and $\sim$44\,d signals.  This diagnostic tests for coherent signals in the RV time series by 
computing the increase in Lomb-Scargle periodogram power near a trial period as additional measurements are taken.  
The 11.4\,d signal rises nearly monotonically in the time series, suggesting a dynamical origin and supporting the planetary interpretation.  
In contrast, the $\sim$44\,d signal waxes and wanes as additional measurements are added to the time 
series, suggesting an incoherent source such as rotationally-modulated spots.

\section{Summary and Discussion}
\label{sec:discussion}

We announce the existence of a low-mass planet orbiting the star \plName, an M2 dwarf 3.6\,pc from Earth. 
Gl\,15\,Ab has a minimum mass \plMsini\,$M_\earth$ and orbits with a period of \plPer\,d.   
The orbital eccentricity is consistent with zero and 
the planet has an equilibrium temperature of 390 K for a Bond albedo $A$ =  0.75 and 550 K for $A$ = 0.  

In addition, we detected a second RV signal with a period of 44~d and an amplitude of 1.8 \mse.   
Because we also detected photometric and chromospheric modulation with the same period, 
we interpret this signal as rotational modulation of spots.  
We will continue to monitor \plName\ to verify that the second signal is truly incoherent, 
as spots should be, and to hunt for additional low-mass planets.  
Our analysis of this second signal shows how challenging the RV detection of Earth-mass planets 
in the habitable zones of early M dwarfs will be.  
Such planets will have periods of $\sim$1--2 months and Doppler amplitudes of $\sim$0.3--0.4 \mse, 
i.e. with similar periods to the spot signal above, but five times smaller in amplitude.  
Detecting such signals will be quite challenging, 
but may be feasible for high-precision Doppler spectrometers with nearly nightly observational coverage.
The high cadence not only improves the (na\"ive) sensitivity as $\sqrt{N_\mathrm{obs}}$, 
it allows the observer to trace out the anomalous RV signature of stellar spots 
and potentially model and subtract it \citep[e.g.][]{Dumusque2012}.
Contemporaneous photometry will also aid in the false positive vetting for such future searches.

Statistical studies of the \textit{Kepler} planet catalog suggest that small planets like \plName b are abundant
\citep{Howard2012,Howard2013} and that multi planet systems are common \citep{Lissauer2011,Fang2012}.  
We plan continued RV monitoring to search for such additional planets in this system.  
Given the distance of only 3.6 pc, we urge high-contrast imaging and astrometry by next generation surveys.  

\acknowledgments{We thank the many observers who contributed to the measurements reported here.  
We gratefully acknowledge the efforts and dedication of the Keck Observatory staff, 
especially Scott Dahm, Greg Doppman, Hien Tran, and Grant Hill for support of HIRES 
and Greg Wirth for support of remote observing.  
We thank Kevin Apps, Andrew Mann, Evan Sinukoff, and Calla Howard for helpful discussions. 
We are grateful to the time assignment committees of the University of Hawaii, the University of California, and NASA 
for their generous allocations of observing time.  
Without their long-term commitment to RV monitoring, this planet would likely remain unknown.  
We acknowledge R.\ Paul Butler and S.\,S.\ Vogt for many years
of contributing to the data presented here.
A.\,W.\,H.\ acknowledges NASA grant NNX12AJ23G.  
G.\,W.\,H.\ acknowledges support from NASA, NSF, Tennessee State University, and
the State of Tennessee through its Centers of Excellence program.
J.\,A.\,J.\ gratefully acknowledges support from generous grants from the
David \& Lucile Packard and Alfred P.\ Sloan Foundations. 
This work made use of the SIMBAD database (operated at CDS, Strasbourg, France), 
NASA's Astrophysics Data System Bibliographic Services, 
and the NASA Star and Exoplanet Database (NStED).
Finally, the authors wish to extend special thanks to those of Hawai`ian ancestry 
on whose sacred mountain of Mauna Kea we are privileged to be guests.  
Without their generous hospitality, the Keck observations presented herein
would not have been possible.}


\pagebreak
\LongTables  

\begin{deluxetable}{cccc}
\tabletypesize{\footnotesize}
\tablecaption{Relative Radial Velocities for Gl 15 A
\label{tab:keck_vels_gl15a}}
\tablewidth{0pt}
\tablehead{
\colhead{}         & \colhead{Radial Velocity}     & \colhead{Uncertainty}  & \colhead{   }  \\
\colhead{BJD -- 2,440,000}   & \colhead{(\mse)}  & \colhead{(\mse)}  & \colhead{$S_{\mathrm{HK}}$}  
}
\startdata
 10461.77113 &   -2.79 &    1.09  &         0.6000                     \\ 
 10716.03838 &   -3.68 &    1.09  &         0.6270                     \\ 
 11044.04091 &    3.60 &    0.81  &         0.6650                     \\ 
 11071.01567 &   -5.12 &    1.07  &         0.5980                     \\ 
 11368.04395 &   -4.06 &    1.04  &         0.5700                     \\ 
 11412.00051 &    2.25 &    1.15  &         0.5690                     \\ 
 11438.83506 &    1.73 &    1.23  &         0.7620                     \\ 
 11552.75997 &   -6.53 &    1.21  &         0.5040                     \\ 
 11704.12495 &    1.26 &    1.06  &         0.5640                     \\ 
 11882.73441 &    2.56 &    1.18  &         0.6340                     \\ 
 12063.12014 &    3.31 &    1.40  &         0.6000                     \\ 
 12097.04192 &   -2.04 &    1.35  &         0.5680                     \\ 
 12099.08799 &    1.79 &    1.25  &         0.5660                     \\ 
 12129.01691 &    1.65 &    1.32  &         0.5630                     \\ 
 12133.09619 &    2.10 &    1.14  &         0.5500                     \\ 
 12487.96446 &    1.27 &    1.20  &         0.5850                     \\ 
 12535.95822 &   -1.55 &    1.38  &         0.5760                     \\ 
 12574.86983 &    2.00 &    1.32  &         0.5310                     \\ 
 12829.10405 &   -1.15 &    1.35  &         0.5360                     \\ 
 12924.89439 &   -2.33 &    1.28  &         \nodata                    \\ 
 13238.98277 &    3.49 &    0.84  &         0.5840                     \\ 
 13302.80391 &   -1.46 &    0.93  &         0.4880                     \\ 
 13338.81208 &   -2.33 &    1.03  &         0.5460                     \\ 
 13547.11252 &    3.52 &    0.57  &         0.5400                     \\ 
 13548.11449 &    3.22 &    0.92  &         0.5300                     \\ 
 13549.12933 &    4.55 &    0.53  &         0.5280                     \\ 
 13550.12107 &    3.65 &    0.61  &         0.5330                     \\ 
 13551.10142 &    3.16 &    0.58  &         0.5575                     \\ 
 13552.06885 &    1.68 &    0.62  &         0.5230                     \\ 
 13571.07965 &    2.24 &    0.54  &         0.5260                     \\ 
 13723.72563 &   -1.74 &    0.71  &         0.5385                     \\ 
 13928.02017 &    2.56 &    0.55  &         0.5720                     \\ 
 13981.94154 &    1.71 &    0.62  &         0.5480                     \\ 
 14085.85430 &    0.06 &    0.80  &         0.5255                     \\ 
 14339.08172 &    2.78 &    0.88  &         0.6430                     \\ 
 14340.04292 &    3.31 &    0.95  &         0.6390                     \\ 
 14398.90264 &   -1.05 &    1.09  &         0.5990                     \\ 
 14429.88240 &    3.59 &    1.02  &         0.6160                     \\ 
 14667.99675 &   -1.91 &    0.96  &         0.5600                     \\ 
 14672.02334 &   -0.82 &    0.74  &         0.5255                     \\ 
 14673.00572 &   -2.46 &    0.61  &         0.5175                     \\ 
 14674.11984 &    0.12 &    0.89  &         0.5420                     \\ 
 14676.09669 &   -4.85 &    0.68  &         0.5220                     \\ 
 14689.13119 &   -0.39 &    0.98  &         0.6190                     \\ 
 14690.09869 &    1.04 &    0.95  &         0.5370                     \\ 
 14721.03229 &   -0.81 &    1.05  &         0.5550                     \\ 
 14778.86564 &   -3.02 &    1.09  &         0.4860                     \\ 
 14807.88673 &    3.02 &    1.05  &         0.5410                     \\ 
 15134.90869 &   -4.21 &    1.02  &         0.5290                     \\ 
 15412.09294 &    2.11 &    0.96  &         0.4720                     \\ 
 15434.05192 &   -1.27 &    0.90  &         0.5200                     \\ 
 15435.03590 &   -2.48 &    0.82  &         0.4940                     \\ 
 15436.05321 &    0.29 &    0.97  &         0.4860                     \\ 
 15437.07327 &    0.31 &    0.99  &         0.5260                     \\ 
 15470.07431 &    2.11 &    1.03  &         0.5210                     \\ 
 15490.02622 &   -3.35 &    1.11  &         0.4420                     \\ 
 15528.86024 &    2.60 &    1.23  &         0.4850                     \\ 
 15542.83596 &    2.25 &    1.02  &         0.5070                     \\ 
 15545.79628 &   -2.97 &    1.08  &         0.4880                     \\ 
 15584.73104 &    1.83 &    1.13  &         0.5180                     \\ 
 15613.71949 &   -1.79 &    1.12  &         0.4840                     \\ 
 15704.11022 &   -1.82 &    1.15  &         0.5090                     \\ 
 15705.11833 &   -5.05 &    1.02  &         0.5210                     \\ 
 15706.11861 &   -4.46 &    0.97  &         0.5110                     \\ 
 15723.10704 &    4.86 &    0.57  &         0.4863                     \\ 
 15726.12358 &    6.12 &    0.57  &         0.4990                     \\ 
 15727.10118 &    3.55 &    0.58  &         0.5210                     \\ 
 15729.07256 &    1.41 &    0.79  &         0.5375                     \\ 
 15731.08821 &    3.07 &    0.53  &         0.5337                     \\ 
 15732.11181 &    2.33 &    0.95  &         0.5420                     \\ 
 15734.10300 &    3.56 &    0.55  &         0.5540                     \\ 
 15735.10478 &    3.93 &    0.54  &         0.5353                     \\ 
 15736.09232 &    3.27 &    0.56  &         0.5317                     \\ 
 15752.12244 &   -6.05 &    0.59  &         0.4737                     \\ 
 15753.02718 &   -5.65 &    0.64  &         0.4593                     \\ 
 15760.12146 &   -3.03 &    0.61  &         0.4450                     \\ 
 15761.11957 &   -3.86 &    0.54  &         0.4463                     \\ 
 15762.12115 &   -2.80 &    0.50  &         0.4370                     \\ 
 15770.11306 &    3.10 &    0.60  &         0.4967                     \\ 
 15770.94986 &    2.66 &    0.52  &         0.5213                     \\ 
 15782.09409 &    1.67 &    0.54  &         0.5440                     \\ 
 15783.10066 &    1.74 &    0.59  &         0.5480                     \\ 
 15786.12265 &   -0.05 &    0.55  &         0.5343                     \\ 
 15787.99817 &    0.10 &    0.58  &         0.5217                     \\ 
 15789.09994 &    1.76 &    0.52  &         0.5203                     \\ 
 15790.09912 &    2.32 &    0.53  &         0.5370                     \\ 
 15791.11511 &    4.01 &    0.55  &         0.5150                     \\ 
 15792.08892 &    3.76 &    0.51  &         0.5070                     \\ 
 15793.09169 &    2.52 &    0.52  &         0.4887                     \\ 
 15794.13095 &    2.12 &    0.53  &         0.4933                     \\ 
 15795.11032 &   -1.26 &    0.58  &         0.4870                     \\ 
 15796.09957 &   -2.35 &    0.56  &         0.4857                     \\ 
 15797.11759 &   -4.85 &    0.57  &         0.4880                     \\ 
 15798.10397 &   -4.62 &    0.57  &         0.5193                     \\ 
 15799.08373 &   -2.81 &    0.56  &         0.4947                     \\ 
 15806.83968 &   -1.14 &    0.64  &         0.4497                     \\ 
 15808.06252 &   -0.07 &    0.66  &         0.4370                     \\ 
 15809.03505 &    0.67 &    0.63  &         0.4497                     \\ 
 15810.08551 &    0.87 &    0.57  &         0.4433                     \\ 
 15811.04909 &   -0.61 &    0.57  &         0.4670                     \\ 
 15812.06251 &    0.20 &    0.53  &         0.4620                     \\ 
 15815.14191 &    5.98 &    0.59  &         0.4940                     \\ 
 15841.84563 &   -4.05 &    0.62  &         0.4930                     \\ 
 15842.84489 &   -2.81 &    0.65  &         0.4730                     \\ 
 15843.93703 &   -2.74 &    0.61  &         0.4727                     \\ 
 15850.94268 &    0.30 &    0.62  &         0.4707                     \\ 
 15851.83105 &   -0.44 &    0.58  &         0.4373                     \\ 
 15853.76621 &   -1.94 &    0.58  &         0.4367                     \\ 
 15870.93426 &    1.28 &    0.74  &         0.5607                     \\ 
 15877.89696 &   -4.58 &    0.65  &         0.5207                     \\ 
 15878.86899 &   -3.80 &    0.61  &         0.5217                     \\ 
 15879.94263 &   -3.82 &    0.59  &         0.5037                     \\ 
 15880.85881 &   -1.00 &    0.57  &         0.5250                     \\ 
 15901.91394 &   -0.77 &    0.57  &         0.4713                     \\ 
 15902.79143 &   -0.71 &    1.48  &         0.4940                     \\ 
 15903.76626 &    0.92 &    0.56  &         0.4703                     \\ 
 15904.81574 &    2.57 &    0.63  &         0.4743                     \\ 

\enddata
\end{deluxetable}

\pagebreak

\bibliographystyle{apj}
\bibliography{gl15a}

\enddocument